\documentclass[
aps,prd,
12pt,
nopreprintnumbers,
showpacs,
eqsecnum,
nofootinbib
]{revtex4-1}

\usepackage{graphicx}
\usepackage{amssymb}

\def\Tr{{\rm Tr\,}}

\begin{document}

\title{Third quantization for scalar and spinor wave functions of the Universe in
an extended minisuperspace}
\author{Nahomi Kan}\email[]{kan@gifu-nct.ac.jp}
\affiliation{National Institute of Technology, Gifu College,
Motosu-shi, Gifu 501-0495, Japan}
\author{Takuma Aoyama}\email[]{b014vbv@yamaguchi-u.ac.jp}
\affiliation{
Graduate School of Sciences and Technology for Innovation, Yamaguchi
University, Yamaguchi-shi, Yamaguchi 753--8512, Japan}
\author{Taiga Hasegawa}\email[]{a019vbu@yamaguchi-u.ac.jp}
\affiliation{
Graduate School of Sciences and Technology for Innovation, Yamaguchi
University, Yamaguchi-shi, Yamaguchi 753--8512, Japan}
\author{Kiyoshi Shiraishi}\email[]{shiraish@yamaguchi-u.ac.jp}
\affiliation{
Graduate School of Sciences and Technology for Innovation, Yamaguchi
University, Yamaguchi-shi, Yamaguchi 753--8512, Japan}
\date{\today}

\begin{abstract}
We consider the third quantization in quantum cosmology of a minisuperspace
extended by the Eisenhart--Duval lift. We study the third quantization based on
both Klein--Gordon type and Dirac-type equations in the extended minisuperspace.
Spontaneous creation of ``universes'' is investigated upon the quantization of a
simple model. We find that the quantization of the  Dirac-type wave function
reveals that the number density of universes is expressed by the Fermi--Dirac
distribution. We also calculate the entanglement entropy of the multi-universe
system.
\end{abstract}



\maketitle

\section{Introduction}
\label{introduction}

There are many problems with the Wheeler--DeWitt (WDW) equation in quantum
cosmology \cite{HH,Hawking,Halliwell,Kiefer0,Kiefer1}. The WDW equation is a
Klein--Gordon-type second-order differential equation even in
minisuperspace.
The Klein--Gordon system has no positive-definite and conserved quantity,
as we learned in the first class of the field theory.
To obtain a conserved current and a positive-definite probability density, a
square root formulation of the WDW equation was proposed originally by authors
of Refs.~\cite{RS,Ryan} and further studied by authors of
Refs.~\cite{DHO,KO,SC,YH,HA,RAH}.%
\footnote{The Dirac-like structures naturally appear also
in quantum cosmology of supergravities \cite{Moniz,Moniz2,Moniz3,Moniz4}.} 
Incidentally, some kind of square root formulation can be found in the
factorization that appears in supersymmetric quantum mechanics. The application
of such elegant and mathematical structures to quantum cosmolgy is also being
enthusiastically studied by authors of Ref.~\cite{JRM}.

In quantum field theory, the wave equation is interpreted not regarding the wave
function as a state vector, but  regarding it as a second quantized field.
The quantized field is expanded with creation and annihilation operators. The
vacuum state is the one that becomes zero by applying the annihilation operator,
and the particles are created by applying the creation operators.
The conserved current is reinterpreted as the one associated with  the charge,
which is defined as positive for particles and negative for antiparticles.

Therefore, we can interpret the wave function of the Universe as a quantum field
in minisuperspace, and the field of the universe is accompanied with the state of 
many universes. Such an approach has been called the third quantization
\cite{CM,McGuigan1,McGuigan2,Strominger,OEF1,OEF2,Perez,HM,Ohkuwa,SC,PM,
Kim,Campanelli,BGMU,Gasperini}.

In the previous paper \cite{KAHS}, the present authors considered a
cosmology of a homogeneous and isotropic space that contains a gravitating
minimally coupled scalar field, and extended the corresponding minisuperspace
description with an additional degree of freedom by a method of the
Eisenhart--Duval lift
\cite{Eisenhart,Duval,DGH,CPC,Pettini,Cariglia,CGGH,FK,Finn,DSS}. The
Eisenhart--Duval lift is one of the classical methods in a Hamiltonian dynamical
system, which allows for a geometric description of the system even in the
presence of the potential term.  We investigated the WDW
equation in quantum cosmology obtained from the
Hamiltonian constraint of the system, by requiring covariance in the extended
minisuperspace. Using the covariance in the extended minisuperspace, we also
constructed an associated Dirac equation for spinor
wave function of the Universe. The introduction of the
Dirac-type equation is originally motivated for
obtaining a positive-definite probability density \cite{DHO,KO,SC,YH,HA,RAH}. We
found fundamental solutions to the WDW equation ($\nabla^2\Phi=0$) and Dirac
equations ($D\!\!\!\!/\,\Psi=0$) in the extended minisuperspace of some simple
models, of which scalar curvatures are zero. The analysis solves some
factor-ordering problems
\cite{HP,Moss,Halliwell2,KW} in quantum cosmology.

Since the equations obtained in the previous paper
are conventional forms in use of the Laplacian operator and the Dirac operator
which appear in general field theories, it is natural to bring them to the third
quantization scheme
\cite{CM,McGuigan1,McGuigan2,Strominger,OEF1,OEF2,Perez,HM,Ohkuwa,SC,PM,
Kim,Campanelli}. So far, the third quantization of the Dirac-type wave function
of the Universe has not been studied.  

In the present paper, we consider the third quantization of quantum
cosmology in the extended minisuperspace. We consider the simplest case of a
homogeneous and isotropic  Friedmann-Lema\^{\i}tre--Robertson--Walker (FLRW)
universe containing a single spatially constant scalar field. In
Section~\ref{sec2}, we briefly review
 the extension of the minisuperspace by the Eisenhart--Duval lift  of the WDW
equation, studied in Ref.~\cite{KAHS}. We show introduction of a
fictitious additional degrees of freedom in the simplest model. The
Klein--Gordon-type and the Dirac-type equations in the extended minisuperspace
are constructed. In Section~\ref{sec3}, we study the third quantization in terms
of two types of equations. The production of bosonic and fermionic universes
from ``nothing'' are discussed and the distribution is calculated. We attempt to
consider the extra degree of freedom as a ``real coordinate'' of the extended
minisuperspace in Section~\ref{sec4}.  The entanglement entropy of universes is
calculated in Section~\ref{sec5}. The last section will be devoted to discussions
and future prospects.

\section{A lightning course of Eisenhart--Duval lift for the WDW equation}
\label{sec2}

In this section, we review an application of Eisenhart--Duval lift to
minisuperspace quantum cosmology.
For details, please consult our previous paper \cite{KAHS}.

As one of the simplest examples, we specify the action of the
gravitating scalar field $\phi$ with a positive cosmological constant $V$:
\begin{equation}
S=\int d^4x\sqrt{-g}\left[\frac{1}{12}R-\frac{1}{2}g^{\mu\nu}
\partial_\mu\phi\partial_\nu\phi-V\right]\,,
\label{action1}
\end{equation}
where $g$ is the determinant of the metric tensor
$g_{\mu\nu}$,
$R$ denotes the scalar curvature constructed from $g_{\mu\nu}$ $(\mu,\nu=0,1,2,3)$,
and  $g^{\mu\nu}$ means the inverse of $g_{\mu\nu}$. Here we use the units
$\hbar=c=4\pi G/3=1$, where $G$ is Newton's constant.  

As the metric, we assume the FLRW metric with a flat space,
\begin{equation}
g_{\mu\nu}dx^\mu dx^\nu=-dt^2+a^2(t)d\mathbf{x}^2\,,
\end{equation}
where $a$ represents the scale factor, and assume that the scalar field depends
only on time
$t$.

The WDW equation
\begin{equation}
\hat{\cal H}\Phi=0\,
\end{equation}
is the quantum Hamiltonian constraint, where
$\Phi$ denotes the wave function of the Universe.
In the present system, it is known that the WDW equation in the minisuperspace
presented by
$x$ and
$y$ takes the form
\begin{equation}
\left(\frac{\partial^2}{\partial x^2}-
\frac{\partial^2}{\partial y^2}
+2Ve^{2nx}\right)\Phi(x,y)=0\,,
\label{bWDW}
\end{equation}
where $n=3$, $x=\ln a$ and $y=\phi$. In this equation, the spatial volume is
interpreted to be appropriately normalized. Here we use the symbol $n$
instead of the fixed number
$3$, because the WDW equations for other some simple models are known to be
written in the similar form.%
\footnote{For example, there are the models treated in
Refs.\cite{Kiefer0,HP2,Kiefer2} ($n=2$), Ref.~\cite{ALNW}
($n=3$), and Ref.~\cite{Paliathanasis} ($n$ depends on values of parameters).}

The essential idea of Eisenhart--Duval lift is to introduce the metric 
of the extended space with an auxiliary dimension $z$, that is, in the present
case,
\begin{equation}
G_{MN}dX^MdX^N={2Ve^{2nx}}(-dx^2+dy^2)+dz^2\,,
\end{equation}
where $X^M=(x,y,z)$, and
replace the WDW equation by
the Laplace equation in the extended
minisuperspace:
\begin{equation}
\frac{1}{\sqrt{-G}}\partial_M(\sqrt{-G}G^{MN}\partial_N\Phi)=\left[
\frac{1}{2Ve^{2nx}}\left(-\frac{\partial^2}{\partial
x^2}+\frac{\partial^2}{\partial y^2}
\right)+\frac{\partial^2}{\partial z^2}\right]\Phi=0\,,
\end{equation}
Here, $G^{MN}$ is the inverse of $G_{MN}$, $G=-(2V)^2e^{4nx}$ is the determinant
of $G_{MN}$, the derivatives are expressed as
$\partial_M\equiv\frac{\partial}{\partial X^M}$.
If we impose an additional constraint
\begin{equation}
\frac{\partial^2\Phi}{\partial z^2}=-\Phi\,,
\label{2c}
\end{equation}
we can reproduce the conventional WDW equation (\ref{bWDW}).%
\footnote{Note that the constraint (\ref{2c}) can be satisfied if we simply take
$\Phi(x,y,z)=
\Phi(x,y) e^{iz}$ concretely.}

Incidentally, the scalar curvature ${\cal R}$ constructed
from
$G_{MN}$, which is defined by
\begin{equation}
{\cal
R}=G^{MN}\left(\partial_{L}\Gamma^{L}_{MN}-\partial_{M}\Gamma^{L}_{NL}
+\Gamma^{L}_{MN}\Gamma^{P}_{LP}-\Gamma^{L}_{MP}\Gamma^{P}_{NL}\right)\,,
\end{equation}
with the Christoffel symbol $\Gamma^L_{MN}$
\begin{equation}
\Gamma^L_{MN}=\frac{1}{2}G^{LP}(\partial_MG_{PN}+\partial_NG_{PM}
-\partial_PG_{MN})\,,
\end{equation}
vanishes in the present case.

The two fundamental solutions of the WDW equation are known to be
\begin{equation}
\phi^{(1)}_\nu(x,y)\propto J_{-i\nu/n}(\sqrt{2V}e^{nx}/n)e^{i\nu y}\,,
\end{equation}
\begin{equation}
\phi^{(2)}_{\nu}(x,y)\propto J_{i\nu/n}(\sqrt{2V}e^{nx}/n)e^{i\nu y}\,,
\end{equation}
where the functions $J_\nu(Z)$ is the
Bessel function of order $\nu$ \cite{Bateman}.


The idea of square-rooting the WDW equation can be found in
Refs.~\cite{DHO,KO,SC,YH,HA,RAH} and others. The Dirac equation in the extended
minisuperspace is fixed in a unique form. It is notable that the Dirac equation
(without mass term) has conformal covariance. 

The Dirac-like equation for a spinor wave function $\Psi$ in the extended
minisuperspace can be written down as
\begin{equation}
D\!\!\!\!/\,\Psi\equiv\hat{\gamma}^MD_M\Psi\equiv\gamma^A e_A^MD_M\Psi=0\,.
\label{DE}
\end{equation}
Here, the constant gamma matrices in the flat spacetime $\gamma^A$ ($A=1,2,3$) are
$\gamma^1=\sigma^1$, $\gamma^2=i\sigma^2$, and $\gamma^3=i\sigma^3$,
where $\sigma^1$, $\sigma^2$, $\sigma^3$ are the Pauli matrices. Note that
$\{\gamma^A,
\gamma^B\}=-2\eta^{AB}$, where $\eta^{AB}=\eta_{AB}=\mbox{diag.}(-1,1,1)$.
The dreibein $e^A_M=\mbox{diag.} ((2V)^{1/2}e^{nx}, (2V)^{1/2}e^{nx},
1)$ is defined through $\eta_{AB}e^A_Me^B_N=G_{MN}$,
and $e_A^M=\mbox{diag.} ((2V)^{-1/2}e^{-nx}, (2V)^{-1/2}e^{-nx},
1)$ is its inverse matrix. Subsequently, we find that $\{\hat{\gamma}^M,
\hat{\gamma}^N\}=-2G^{MN}$. The covariant derivative $D_M$ for the spin connection
$\omega_{MAB}$ is defined as
$D_M\equiv\partial_M+\frac{1}{4}
\omega_{MAB}\Sigma^{AB}$, where $\Sigma^{AB}\equiv-\frac{1}{2}[\gamma^A,
\gamma^B]$. The spin connection $\omega_{MAB}$ is given by
\begin{equation}
\omega_{MAB}=\frac{1}{2}e^N_A(\partial_Me_{NB}-\Gamma^L_{MN}e_{LB})
-(A\leftrightarrow B)\,.
\end{equation}
In the present model, one can find $\omega_{y12}=-\omega_{y21}=-n$.

In the extended minisuperspace presently considered, we find that the Dirac
equation (\ref{DE}) is equivalent to
\begin{equation}
\left[\sigma^1\left(\frac{\partial}{\partial x}+\frac{n}{2}\right)+
i\sigma^2\frac{\partial}{\partial y}+
i\sigma^3\sqrt{2V}e^{nx}\frac{\partial}{\partial z}\right]\Psi=0\,.
\end{equation}
In order to reduce the equation to that of physical variables $x$ and $y$,
we choose the additional constraint on $\Psi$:
\begin{equation}
\frac{1}{i}\frac{\partial}{\partial z}\Psi=\Psi\,.
\end{equation}
Now, the Dirac equation reads in the matrix form, 
\begin{equation}
\left(
\begin{array}{cc}
-\sqrt{2V}e^{n x} &
\frac{\partial}{\partial
x}+\frac{n}{2}+\frac{\partial}{\partial y}\\
\frac{\partial}{\partial
x}+\frac{n}{2}-\frac{\partial}{\partial y} &
\sqrt{2V}e^{n x}
\end{array}
\right)\left(
\begin{array}{c}
\Psi_{+} \\ \Psi_{-}
\end{array}
\right)=\left(
\begin{array}{c}
0 \\ 0
\end{array}
\right)\,,
\label{DE1}
\end{equation}
where $\Psi={\Psi_{+}\choose\Psi_{-}}$.
The fundamental solutions of the equation are found to be
\begin{equation}
\psi^{(1)}_{\pm,\nu}\propto\pm
J_{-\frac{i\nu}{n}\pm\frac{1}{2}}(\sqrt{2V}e^{nx}/n)e^{i\nu
y}\,,\quad\psi^{(2)}_{\pm,\nu}\propto
J_{\frac{i\nu}{n}\mp\frac{1}{2}}(\sqrt{2V}e^{nx}/n)e^{i\nu
y}\,.
\end{equation}

\section{Third quantization and spontaneous creation of multiple universes}
\label{sec3}

The third quantization is attained by taking the wave function for an
operator called as a ``field'' acting on the state vectors of a system of multiple
universes. First we review the result of Hosoya and Morikawa \cite{HM}
for the scalar wave function of the conventional WDW equation in the present
model.

We start with the wave function $\Phi(x,y)$ that satisfies the WDW equation
(\ref{bWDW}).
The WDW equation is a hyperbolic second-order differential
equation, and here we regard $x$ as the intrinsic time coordinate in
minisuperspace. Note that when $x$ varies from $-\infty$ to $\infty$, the scale
factor $a=e^{x}$ grows from $0$ to $+\infty$.

We expand the real scalar function $\Phi(x,y)$ as
\begin{equation}
\Phi(x,y)
=\int^\infty_{-\infty} d\nu\, \Bigl[a_\nu \phi_\nu(x,y)+a^\dagger_\nu
\phi^*_\nu(x,y)\Bigr]\,,
\end{equation}
where $\phi_\nu$ and $\phi^*_\nu$ are the normalized fundamental solutions of the
WDW equation (\ref{bWDW}) and satisfy the orthonormal condition
\begin{equation}
(\phi_\nu,\phi_{\nu'})=\delta(\nu-\nu')\,,\quad
(\phi^*_\nu,\phi_{\nu'})=0\,,
\end{equation}
where the inner product is defined as
\begin{equation}
(f,g)\equiv i\int dy \sqrt{-G}|G^{xx}|
(f^*\partial_xg-\partial_xf^*g)=i\int dy
(f^*\partial_xg-\partial_xf^*g)
\,.
\end{equation}
Note that $(g,f)=(f,g)^*=-(f^*,g^*)$.
Since this prescription is analogous to the one in quantum field theory in curved
spacetime \cite{BD}, $a_\nu$ and $a^\dagger_\nu$ are corresponding to annihilation
and creation operators, respectively.
We assume the following bosonic commutation relations
\begin{equation}
[a_\nu, a_{\nu'}^\dagger]=\delta(\nu-\nu')\,,\,
[a_\nu, a_{\nu'}]=[a_\nu^\dagger, a_{\nu'}^\dagger]=0
\quad
\Leftrightarrow\quad
[\Phi(x,y), \partial_x\Phi(x,y')]=i\delta(y-y')\,,
\end{equation}
which realize the canonical commutation relation of a scalar field theory.
Hence the vacuum state $|0\rangle$ is defined by
\begin{equation}
a_\nu|0\rangle=0\quad\mbox{(for all $\nu$)}\,,
\end{equation}
and normalized as $\langle0|0\rangle=0$. The Fock space of this ``bosonic''
multiple universes is spanned by the states such as
$a^\dagger_{\nu_1}a^\dagger_{\nu_2}\cdots|0\rangle$, etc.

According to  quantum field theory in curved space \cite{BD}, the vacuum state
is not unique but depends  on the different region in spacetime.
In the present case, we define in and out regions in minisuperspace for
$x\rightarrow-\infty$ (when the scale factor goes to zero) and
$x\rightarrow\infty$ (when the scale factor grows infinitely), respectively.
 When $x\rightarrow-\infty$, $J_{-i|\nu|/n}(\sqrt{2V}e^{nx}/n)$
is proportional to 
$e^{-i|\nu|x}$ \cite{Bateman}. Thus, the positive-frequency, in-mode function can
be written as
\begin{equation}
\phi^{in}_\nu(x,y)=
\frac{1}{2}\frac{1}{\sqrt{n\sinh\frac{|\nu|\pi}{n}}}J_{-i|\nu|/n}
(\sqrt{2V}e^{nx}/n)e^{i\nu
y}\,.
\end{equation}

On the other hand, when $x\rightarrow+\infty$, the function
$H^{(2)}_{-i|\nu|/n}(\sqrt{2V}e^{nx}/n)$ is proportional 
$\exp[-i\sqrt{2V}e^{nx}/n]$,
where $H^{(2)}_\nu(Z)$ is the Hankel function of the second kind defined by
\cite{Bateman}
\begin{equation}
H^{(2)}_{\nu}(z)=\frac{-i}{\sin\pi\nu}\left[e^{i\pi\nu}J_\nu(z)-J_{-\nu}(z)\right]\,.
\end{equation}
Thus, the positive-frequency, out-mode
function can be written as
\begin{equation}
\phi^{out}_\nu(x,y)=
\frac{e^{-\frac{\pi|\nu|}{2n}}}{2\sqrt{2n}}H^{(2)}_{-i|\nu|/n}(\sqrt{2V}e^{nx}/n)e^{i\nu
y}\,.
\end{equation}

Now, the field operator $\Phi$ is expanded by two ways:
\begin{eqnarray}
\Phi(x,y)&=&\int^\infty_{-\infty} d\nu [a^{in}_\nu
\phi^{in}_\nu(x,y)+a^{in\dagger}_\nu
\phi^{in*}_\nu(x,y)]\nonumber \\
&=&\int^\infty_{-\infty} d\nu [a^{out}_\nu \phi^{out}_\nu(x,y)+a^{out\dagger}_\nu
\phi^{out*}_\nu(x,y)]\,.
\label{twoway}
\end{eqnarray}
Then, we define the two vacuum states, the in-vacuum $|0,in\rangle=\prod_\nu
|0,in\rangle_\nu$
and the out-vacuum $|0,out\rangle=\prod_\nu|0,out\rangle_\nu$. They satisfy
\begin{equation}
 a^{in}_\nu|0,in\rangle=0\,,\quad a^{out}_\nu|0,out\rangle=0\, \quad
\mbox{(for all $\nu$)}.
\end{equation}
Thus, the in-vacuum state represents no universe, or ``nothing''
in the region $x\rightarrow-\infty$.

From Eq.~(\ref{twoway}), we find
\begin{equation}
a^{out}_\nu=(\phi^{out},\Phi)=\frac{1}{\sqrt{1-e^{-2\pi|\nu|/n}}}a^{in}_\nu
+\frac{1}{\sqrt{e^{2\pi|\nu|/n}-1}}a^{in\dagger}_{-\nu}\,.
\label{oi}
\end{equation}
From this, we can calculate the average number of ``bosonic'' universes $N_{B\nu}$
for the specific value of $\nu$ in the out region $x\rightarrow\infty$, created
from nothing
\begin{equation}
N_{B\nu}\equiv\langle 0,in|a^{out\dagger}_\nu a^{out}_\nu|0,in\rangle
=\frac{1}{e^{2\pi|\nu|/n}-1}\,,
\end{equation}
which is the Planck distribution. The above results have been shown by
the paper of Hosoya and Morikawa \cite{HM}.


Now, we turn to the third quantization for the spinor wave function obeying the
Dirac equation (\ref{DE1}). The spinor wave function is decomposed as
\begin{equation}
\Psi(x,y)=\int^\infty_0 d\nu\, \Bigl[b_\nu u_\nu(x,y)+d^\dagger_{\nu}
u_{-\nu}(x,y)+ b_{-\nu} v_{-\nu}(x,y)+d^\dagger_{-\nu} v_{\nu}(x,y)\Bigr]\,.
\label{eqD}
\end{equation}
It should be noted that the integration parameter $\nu$ runs over positive
values in this expression.
The inner product of two spinors is defined by
\begin{equation}
(\vartheta,\varphi)\equiv \int dy \sqrt{-G}|G^{xx}|^{1/2}
\vartheta^\dagger\varphi=\sqrt{2V}e^{nx}\int dy\, 
\vartheta^\dagger\varphi
\,.
\end{equation}
Note that $(\varphi,\vartheta)=(\vartheta,\varphi)^*$.
The mode functions in (\ref{eqD}) are normalized as
\begin{equation}
(u_\nu,u_{\nu'})=(v_\nu,v_{\nu'})=\delta(\nu-\nu')
\,,\quad
(u_\nu, v_{\nu'})=(v_\nu, u_{\nu'})=0\,.
\end{equation}

As is known in quantum field theory, anticommutators of the fermionic operators
obey%
\footnote{In the third quantization, we consider the description borrowing the
state and creation and annihilation operators from the quantum field theory. Note
that the signature of the minisuperspace is Lorentzian. It is very interesting
that the spin-statistics theorem does not hold, but it is expected that
difficulties will appear when introducing interactions among multiple universes.}
\begin{eqnarray}
& &\{b_\nu, b_{\nu'}^\dagger\}=\{d_\nu, d_{\nu'}^\dagger\}=\delta(\nu-\nu')\,,\,
(\mbox{the other anticommutators})=0 \nonumber \\
& &\qquad\Leftrightarrow\quad
\{\Psi_{\alpha}(x,y),
\Psi_{\beta}^\dagger(x,y')\}=\delta_{\alpha\beta}\delta(y-y')\,,
\end{eqnarray}
where $\alpha, \beta=+,-$.

As in the scalar case, the spinor field can be expressed as
\begin{eqnarray}
\Psi(x,y)&=&\int^\infty_0 d\nu [b^{in}_\nu u^{in}_\nu(x,y)+d^{in\dagger}_{\nu}
u^{in}_{-\nu}(x,y)+ b^{in}_{-\nu} v^{in}_{-\nu}(x,y)+d^{in\dagger}_{-\nu}
v^{in}_{\nu}(x,y)]\nonumber
\\
&=&\int^\infty_0 d\nu [b^{out}_\nu u^{out}_\nu(x,y)+d^{out\dagger}_{\nu}
u^{out}_{-\nu}(x,y)+ b^{out}_{-\nu} v^{out}_{-\nu}(x,y)+d^{out\dagger}_{-\nu}
v^{out}_{\nu}(x,y)]\,.
\end{eqnarray}
Here the in-mode spinor functions are obtained from
\begin{eqnarray}
u^{in}_\nu(x,y)&=&\frac{1}{2\sqrt{n\cosh\frac{\pi\nu}{n}}}
{J_{-i\frac{\nu}{n}+\frac{1}{2}}
({\sqrt{2V}e^{nx}/n})\choose -J_{-i\frac{\nu}{n}-\frac{1}{2}}
({\sqrt{2V}e^{nx}/n})}e^{i\nu y}\,,\nonumber \\
v^{in}_\nu(x,y)&=&\frac{1}{2\sqrt{n\cosh\frac{\pi\nu}{n}}}
{J_{i\frac{\nu}{n}-\frac{1}{2}}
({\sqrt{2V}e^{nx}/n})\choose J_{i\frac{\nu}{n}+\frac{1}{2}}
({\sqrt{2V}e^{nx}/n})}e^{i\nu y}\,,
\end{eqnarray}
while the out-mode spinor functions are given by
\begin{eqnarray}
& &u^{out}_{\nu}(x,y)=U_\nu(x,y)\,,\quad
u^{out}_{-\nu}(x,y)=-iV_{-\nu}(x,y)\,,\quad(\nu>0)\\
& &v^{out}_{-\nu}(x,y)=-iU_{-\nu}(x,y)\,,\quad
v^{out}_{\nu}(x,y)=V_{\nu}(x,y)\,,\quad(\nu>0)
\end{eqnarray}
where
\begin{eqnarray}
U_\nu(x,y)&=&\frac{e^{-\frac{\pi\nu}{2n}}}{2\sqrt{2n}}
{H^{(2)}_{-i\frac{\nu}{n}+\frac{1}{2}}
({\sqrt{2V}e^{nx}/n})\choose -H^{(2)}_{-i\frac{\nu}{n}-\frac{1}{2}}
({\sqrt{2V}e^{nx}/n})}e^{i\nu y}=
\frac{u^{in}_\nu(x,y)}{\sqrt{1+e^{-2\pi\nu/n}}}+
i\frac{v^{in}_\nu(x,y)}{\sqrt{e^{2\pi\nu/n}+1}}
\,, \\
V_\nu(x,y)&=&\frac{i e^{\frac{\pi\nu}{2n}}}{2\sqrt{2n}}
{H^{(1)}_{-i\frac{\nu}{n}+\frac{1}{2}}
({\sqrt{2V}e^{nx}/n})\choose -H^{(1)}_{-i\frac{\nu}{n}-\frac{1}{2}}
({\sqrt{2V}e^{nx}/n})}e^{i\nu y}=
i\frac{u^{in}_\nu(x,y)}{\sqrt{e^{2\pi\nu/n}+1}}+
\frac{v^{in}_\nu(x,y)}{\sqrt{1+e^{-2\pi\nu/n}}}\,.
\end{eqnarray}

Then, one can find the relation among the in- and out-operators as
\begin{eqnarray}
b^{out}_{\nu}&=&\frac{1}{\sqrt{1+e^{-2\pi|\nu|/n}}}b^{in}_{\nu}-i
\frac{1}{\sqrt{e^{2\pi|\nu|/n}+1}}d^{in\dagger}_{-\nu}\,,\quad(\nu>0)
\label{1}\\
b^{out}_{-\nu}&=&\frac{1}{\sqrt{1+e^{-2\pi|\nu|/n}}}b^{in}_{-\nu}+i
\frac{1}{\sqrt{e^{2\pi|\nu|/n}+1}}d^{in\dagger}_{\nu}\,,\quad(\nu>0)
\label{2}\\
 d^{out\dagger}_{\nu}&=&\frac{1}{\sqrt{1+e^{-2\pi|\nu|/n}}}d^{in\dagger}_{\nu}+i
\frac{1}{\sqrt{e^{2\pi|\nu|/n}+1}}b^{in}_{-\nu}\,,\quad(\nu>0)
\label{3}\\
d^{out\dagger}_{-\nu}&=&\frac{1}{\sqrt{1+e^{-2\pi|\nu|/n}}}d^{in\dagger}_{-\nu}-i
\frac{1}{\sqrt{e^{2\pi|\nu|/n}+1}}b^{in}_{\nu}\,.\quad(\nu>0)
\label{4}
\end{eqnarray}
Accordingly, with the aid of $b^{in}_\nu|0,in\rangle=d^{in}_\nu|0,in\rangle=0$
and so on, we find the average number of ``fermionic'' universes in the form
\begin{equation}
N_{F\nu}\equiv\langle 0,in|b^{out\dagger}_\nu b^{out}_\nu|0,in\rangle
=\frac{1}{e^{2\pi|\nu|/n}+1}\,,\quad(-\infty<\nu<\infty)
\end{equation}
and that of ``anti-universes'' as
\begin{equation}
\bar{N}_{F\nu}\equiv\langle 0,in|d^{out\dagger}_\nu d^{out}_\nu|0,in\rangle
=\frac{1}{e^{2\pi|\nu|/n}+1}=N_{F\nu}\,.\quad(-\infty<\nu<\infty)
\end{equation}

We conclude that 
the finite number density of universes is the Fermi--Dirac distribution
in the context of spontaneous creation of fermionic universes derived from the
third quantization for the Dirac-type equation in quantum cosmology.

\section{Is there dependence on the auxiliary dimension of the extended
minisuperspace?}
\label{sec4}

Until the previous section, we have treated the extra degree of freedom, or the
extra auxiliary dimension $z$ of the extended minisuperspace, as a fictitious one.
That is, owing to the additional constraint, we have considered that wave
functions and also quantized fields are defined in two dimensional minisuperspace
as concerned with the present model.

In this section, we consider the third quantized field theory without the
additional constraint in full extended dimensions, i.e., three dimensional
extended minisuperspace in our model.

Again, we restart with the scalar case.
The equation of motion of the scalar field in the extended minisuperspace of our
model is
\begin{equation}
\left[
\frac{1}{2Ve^{2nx}}\left(-\frac{\partial^2}{\partial
x^2}+\frac{\partial^2}{\partial y^2}
\right)+\frac{\partial^2}{\partial z^2}\right]\Phi(x,y,z)=0\,.
\end{equation}
The expansion of the scalar field in terms of mode functions is given by
\begin{eqnarray}
\Phi(x,y,z)&=&\int^\infty_{-\infty} dk \int^\infty_{-\infty} d\nu\,
\Bigl[a^{in}_{\nu,k}
\phi^{in}_{\nu,k}(x,y,z)+a^{in\dagger}_{\nu,k}
\phi^{in*}_{\nu,k}(x,y,z)\Bigr]\nonumber \\
&=&\int^\infty_{-\infty} dk \int^\infty_{-\infty} d\nu\, \Bigl[a^{out}_{\nu,k}
\phi^{out}_{\nu,k}(x,y,z)+a^{out\dagger}_{\nu,k}
\phi^{out*}_{\nu,k}(x,y,z)\Bigr]\,,
\end{eqnarray}
where
\begin{equation}
\phi^{in}_{\nu,k}(x,y,z)=
\frac{1}{2}\frac{1}{\sqrt{2\pi n\sinh\frac{|\nu|\pi}{n}}}J_{-i|\nu|/n}
(|k|\sqrt{2V}e^{nx}/n)e^{i\nu
y+ikz}\,,
\end{equation}
and
\begin{equation}
\phi^{out}_{\nu,k}(x,y,z)=
\frac{e^{-\frac{\pi|\nu|}{4n}}}{\pi\sqrt{2n}}H^{(2)}_{-i|\nu|/n}(|k|\sqrt{2V}e^{nx}/n)e^{i\nu
y+ikz}\,.
\end{equation}

We presume the commutation relations $[a^{in}_{\nu,k},a^{in\dagger}_{\nu',k'}]=
[a^{out}_{\nu,k},a^{out\dagger}_{\nu',k'}]=\delta(\nu-\nu')\delta(k-k')$, etc.
Similar analysis to one in the previous section leads to the relation between in-
and out-operators:
\begin{equation}
a^{out}_{\nu,k}=\frac{1}{\sqrt{1-e^{-2\pi|\nu|/n}}}a^{in}_{\nu,k}
+\frac{1}{\sqrt{e^{2\pi|\nu|/n}-1}}a^{in\dagger}_{-\nu,k}\,,
\end{equation}
whose form is apparently independent of the index $k$.
Then, the number density is found to be independent of $k$ also in this case:
\begin{equation}
N_{B\nu,k}\equiv\langle 0,in|a^{out\dagger}_{\nu,k} a^{out}_{\nu,k}|0,in\rangle
=\frac{1}{e^{2\pi|\nu|/n}-1}\,.
\end{equation}
Therefore we conclude, at least in the present subject to study,
the variable $z$ can be considered as fictitious and unphysical.%
\footnote{This may be trivial if we have observed that the relation of
in- and out-operators are independent of the constant $\sqrt{2V}$.}  
As discussed in our previous paper \cite{KAHS}, the eigen value $|k|$ can be
absorbed in redefinition in the original dynamical system, and thus this is a
natural result.

Just in case, let us consider the case of the spinor fermionic field.
The solution of the Dirac equation (\ref{DE})
\begin{equation}
\left[\sigma^1\left(\frac{\partial}{\partial x}+\frac{n}{2}\right)+
i\sigma^2\frac{\partial}{\partial y}+
i\sigma^3\sqrt{2V}e^{nx}\frac{\partial}{\partial z}\right]\Psi=0\,,
\end{equation}
can be expanded in modes with a new index $k$ as
\begin{eqnarray}
\Psi(x,y,z)&=&\int^\infty_0 dk\int^\infty_0 d\nu\, \Bigl[b^{in}_{\nu,k}
u^{in}_{\nu,k}(x,y,z)+b^{in}_{\nu,-k}
u^{in}_{\nu,-k}(x,y,z)\nonumber \\
& &\qquad\qquad\qquad+d^{in\dagger}_{\nu,k} u^{in}_{-\nu,k}(x,y,z)
+d^{in\dagger}_{\nu,-k} u^{in}_{-\nu,-k}(x,y,z)\nonumber \\
& &\qquad\qquad\qquad+b^{in}_{-\nu,k} v^{in}_{-\nu,k}(x,y,z)
+b^{in}_{-\nu,-k} v^{in}_{-\nu,-k}(x,y,z)\nonumber \\
& &\qquad\qquad\qquad+d^{in\dagger}_{-\nu,k}v^{in}_{\nu,k}(x,y,z)
+d^{in\dagger}_{-\nu,-k}v^{in}_{\nu,-k}(x,y,z)\Bigr]\,,
\end{eqnarray}
and similar expansion in terms of out-modes.

The construction of mode functions are performed as in the previous section
and here we use in-mode functions
\begin{eqnarray}
u^{in}_{\nu,k}(x,y)&=&\frac{1}{2\sqrt{2\pi
kn\cosh\frac{\pi\nu}{n}}}
{J_{-i\frac{\nu}{n}+\frac{1}{2}}
({k\sqrt{2V}e^{nx}/n})\choose -J_{-i\frac{\nu}{n}-\frac{1}{2}}
({k\sqrt{2V}e^{nx}/n})}e^{i\nu y+ikz}\,,\,(k>0) \\
v^{in}_{\nu,k}(x,y)&=&\frac{1}{2\sqrt{2\pi kn\cosh\frac{\pi\nu}{n}}}
{J_{i\frac{\nu}{n}-\frac{1}{2}}
({k\sqrt{2V}e^{nx}/n})\choose J_{i\frac{\nu}{n}+\frac{1}{2}}
({k\sqrt{2V}e^{nx}/n})}e^{i\nu y+ikz}\,,\,(k>0) \\
u^{in}_{\nu,-k}(x,y)&=&\frac{1}{2\sqrt{2\pi
kn\cosh\frac{\pi\nu}{n}}}
{J_{-i\frac{\nu}{n}+\frac{1}{2}}
({k\sqrt{2V}e^{nx}/n})\choose J_{-i\frac{\nu}{n}-\frac{1}{2}}
({k\sqrt{2V}e^{nx}/n})}e^{i\nu y-ikz}\,,\,(k>0) \\
v^{in}_{\nu,-k}(x,y)&=&\frac{1}{2\sqrt{2\pi kn\cosh\frac{\pi\nu}{n}}}
{J_{i\frac{\nu}{n}-\frac{1}{2}}
({k\sqrt{2V}e^{nx}/n})\choose -J_{i\frac{\nu}{n}+\frac{1}{2}}
({k\sqrt{2V}e^{nx}/n})}e^{i\nu y-ikz}\,,\,(k>0) 
\end{eqnarray}
and out-mode functions are constructed from
\begin{eqnarray}
U_{\nu,\pm k}(x,y,z)&=&
\frac{u^{in}_{\nu,\pm k}(x,y,z)}{\sqrt{1+e^{-2\pi\nu/n}}}+
i\frac{v^{in}_{\nu,\pm k}(x,y,z)}{\sqrt{e^{2\pi\nu/n}+1}}
\,,\quad(k>0) \\
V_{\nu,\pm k}(x,y,z)&=&
i\frac{u^{in}_{\nu,\pm k}(x,y,z)}{\sqrt{e^{2\pi\nu/n}+1}}+
\frac{v^{in}_{\nu,\pm k}(x,y,z)}{\sqrt{1+e^{-2\pi\nu/n}}}\,.\quad(k>0)
\end{eqnarray}

Finally, we obtain the result
\begin{equation}
\langle 0,in|b^{out\dagger}_{\nu,k} b^{out}_{\nu,k}|0,in\rangle
=\langle 0,in|d^{out\dagger}_{\nu,k} d^{out}_{\nu,k}|0,in\rangle
=\frac{1}{e^{2\pi|\nu|/n}+1}\,.\quad(-\infty<\nu, k<\infty)
\end{equation}
We thus confirmed the $k$-independence in the third quantization method
for the case with the fermionic field.

\section{Entanglement entropy of multiple universes}
\label{sec5}
As in a quantum system,
the von Neumann entropy in an expanding universe has been calculated
in many works including Ref.~\cite{BFS,FMMM}.
The von Neumann entropy is an index for quantatively evaluating
entanglement
\cite{HHHH,Nishioka}. 
It has been demonstrated that the von Neumann entropy can be calculated in
third-quantized quantum cosmology
\cite{BM}, and
we will show both bosonic and fermionic system originated from the usual WDW
equation and the Dirac-type equation in our present model. We return to the
constrained version, or we discard the index
$k$, as in Section~\ref{sec3}.

First, we examine ``bosonic'' universes originated from the wave function of the
usual WDW equation. From Eq.~(\ref{twoway}), we find
\begin{equation}
a^{in}_\nu=\frac{1}{\sqrt{1-e^{-2\pi|\nu|/n}}}a^{out}_\nu
-\frac{1}{\sqrt{e^{2\pi|\nu|/n}-1}}a^{out\dagger}_{-\nu}\,.
\label{io}
\end{equation}
For simplicity, we focus on the fixed mode of $\nu$.
If $|0,in\rangle_\nu=\sum_{p=0}^\infty A_p |p_\nu,p_{-\nu},out\rangle$,
where
$|p_\nu,p'_{-\nu},out\rangle=|p,out\rangle_\nu|p',out\rangle_{-\nu}=\frac{1}{\sqrt{p!}}(a_\nu^{out\dagger})^p|0,out\rangle_\nu
\frac{1}{\sqrt{p'!}}(a_{-\nu}^{out\dagger})^{p'}|0,out\rangle_{-\nu}$, is
assumed,
$a^{in}_\nu|0,in\rangle_\nu=0$ with the use of Eq.~(\ref{io}) yields
\begin{equation}
|0,in\rangle_\nu=\sqrt{1-e^{-2\pi|\nu|/n}}\sum_{p=0}^\infty e^{-p\pi|\nu|/n}
|p_\nu,p_{-\nu},out\rangle\,.
\end{equation}

After the modes with the order $-\nu$ are traced out, we obtain the reduced
density matrix 
\begin{equation}
\rho_\nu\equiv\Tr_{-\nu}\Bigl[|0,in\rangle_\nu{}_\nu\langle 0,in|\Bigr]
=(1-e^{-2\pi|\nu|/n})\sum_{p=0}^\infty e^{-2\pi|\nu|p/n}
|p,out\rangle_\nu{}_\nu\langle p,out|\,.
\end{equation}
Using this, the von Neumann entanglement entropy defined by
\begin{equation}
S\equiv-\Tr\Bigl[\rho_\nu\log_2\rho_\nu\Bigr]\,.
\end{equation}
becomes
\begin{eqnarray}
S_B=\log_2\left[
\frac{\exp[{\frac{2\pi|\nu|/n}{e^{2\pi|\nu|/n}-1}}]}{1-e^{-2\pi|\nu|/n}}\right]\,,
\end{eqnarray}
where the subscript ``$B$'' is assigned because universes are bosonic in this
case.

Incidentally, the R\'enyi entropy, which is another index of the entanglement,
is defined as \cite{HHHH,Nishioka}
\begin{equation}
S(q)\equiv\frac{1}{1-q}\log_2\Bigl[\Tr\rho_\nu^q\Bigr]\,.
\end{equation}
Note that the limit of $q\rightarrow 1$ reduces the R\'enyi entropy to the von
Neumann entropy exactly. In our bosonic case, one can find
\begin{eqnarray}
S_{B}(q)=\frac{1}{1-q}\log_2\left[
\frac{(1-e^{-2\pi|\nu|/n})^q}{1-e^{-2\pi|\nu|q/n}}\right]\,.
\end{eqnarray}

Next, we consider the fermionic universes originated from the Dirac field.
We consider an in-vacuum on a specific $\nu$,
$|0,in\rangle_\nu|\bar{0},in\rangle_{-\nu}$, where
$b_\nu^{in}|0,in\rangle_\nu=d_{-\nu}^{in}|\bar{0},in\rangle_{-\nu}=0$.
Similarly to the bosonic case, the inverse relation of
Eq.~(\ref{1}--\ref{4}) leads to
\begin{equation}
|0,in\rangle_\nu|\bar{0},in\rangle_{-\nu}=\frac{1}{\sqrt{1+e^{-2\pi|\nu|/n}}}
\left(|0,out\rangle_\nu|\bar{0},out\rangle_{-\nu}-i\,\mathrm{sgn}(\nu)
e^{-\pi|\nu|/n}|1,out\rangle_\nu|\bar{1},out\rangle_{-\nu}\right)\,,
\end{equation}
where $|1,out\rangle_\nu\equiv b^{out\dagger}_\nu|0,out\rangle_\nu$
and $|\bar{1},out\rangle_{-\nu}\equiv
d^{out\dagger}_{-\nu}|\bar{0},out\rangle_{-\nu}$, of course with the relations
$b_\nu^{out}|0,out\rangle_\nu=d_{-\nu}^{out}|\bar{0},out\rangle_{-\nu}=0$
for the out-vacuum.

Tracing over the anti-universe modes with the index $-\nu$, we obtain
\begin{eqnarray}
\rho_\nu&\equiv&\Tr_{-\nu}\Bigl[|0,in\rangle_\nu|\bar{0},in\rangle_{-\nu}
\langle\bar{0},in|_{-\nu}\langle
0,in|_\nu\Bigr]\nonumber \\
&=&\frac{1}{1+e^{-2\pi|\nu|/n}}|0,out\rangle_\nu\langle
0,out|_\nu+\frac{e^{-2\pi|\nu|/n}}{1+e^{-2\pi|\nu|/n}}  |1,out\rangle_\nu\langle
1,out|_\nu\,.
\end{eqnarray}
From this reduced matrix, we find the von Neumann entropy
\begin{eqnarray}
S_F=\log_2\left(
\frac{1+e^{-2\pi|\nu|/n}}{\exp[{-\frac{2\pi|\nu|/n}{e^{2\pi|\nu|/n}+1}}]}\right)\,,
\end{eqnarray}
and the R\'enyi entropy
\begin{eqnarray}
S_{F}(q)=\frac{1}{1-q}\log_2\left[
\frac{1+e^{-2\pi|\nu|q/n}}{(1+e^{-2\pi|\nu|/n})^q}\right]\,.
\end{eqnarray}

\begin{figure}[ht]
\centering
\includegraphics[width=6cm]
{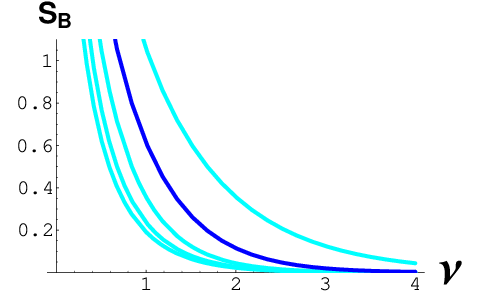}\quad
\includegraphics[width=6cm]
{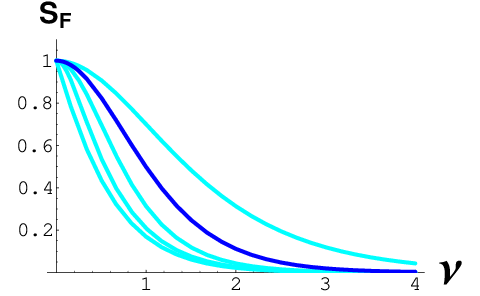}\\
(a) \hspace{6cm} (b) 
\caption{The bosonic case (a) and the fermionic case
(b): The entanglement entropy as a function of $\nu$ for different
values of $q$ ($q=0.5,1,2,5,100$, from the upper line to the lower line,
and only the line for $q=1$ is drawn in blue), with
$n=3$. }
\label{fig1}
\end{figure}

In Fig.~1, the dependence of the entanglement entropy on $\nu$ for different
values of the parameter $q$.
The entanglement monotonically decreases with $\nu$ for both cases, while
it diverges when $\nu\rightarrow 0$ only in the bosonic case.


If the principle of maximum entropy can be applied to the initial state of the
multiple universes, it may be said that the universe starts with a small value of
the mode number $\nu$ of the scalar field. In other words, the probability
distribution of the value of the scalar field will be nearly uniform.


Before closing this section, we must mention the physical meaning of
the entanglement entropy of the multiple-universe system.
The problem of observers already existed in quantum cosmology (who exists
outside the Universe?). 
There is no observer in quantum cosmology, where no measurement was carried out.
Thus, the definition of information or entropy deduced from measurement of a
system should be ambiguous.

Nevertheless, Hosoya and Morikawa \cite{HM} considered ``detector'' of universes,
as an ambitious attempt.
Unfortunately, their approach may bring about more ambiguities in the model,
because it should be additionally assumed the interaction with multiple universes.
The inclusion of interactions, or ``wormholes'' making bridges between
(baby) universes, is still an interesting idea \cite{Strominger}, since the
interaction may fix the physical constants in our (specific) universe.

\section{Discussion and outlook}
\label{conclusion}

In conclusion, we have investigated the third quantization of quantum cosmology
in a simple model by using the extended minisuperspace.
As the equation of motion of the quantized field, we considered both
Klein--Gordon-type and Dirac-type equations, which describe annihilation and
creation of bosonic and fermionic universes, respectively. By utilizing the
technique of quantum field theory in curved space, we obtain the exact results on
the average number of bosonic and fermionic universes which are spontaneously
created. It is found that the average number is expressed by the Planck
distribution for bosonic universes and the Fermi--Dirac distribution for
fermionic universe, respectively.
We also show the calculable entanglement entropies for both systems.
Although we have studied a simple model in this paper, exact results may play some
important role in understanding the crude concepts of the third quantization in
future study. The analyses in various cosmological models are required as a
straightforward extension of the present study.

We have chosen the in-vacuum state as the initial state, but
condensate/degenerate states are equally considerable as a hypothesis.
On the other hand, for operators, interaction terms may fix the constant in nature
\cite{Strominger} and provide significance of entanglement.
Moreover, the dependence of physical quantities on the auxiliary dimension is
quite possible if we consider general interaction of fields.
We can postulate supersymmetric third-quantized field theory as well.
The study in the above mentioned directions will be advanced in future
publication.

\section*{Note added}
After completion of the present manuscript, Prof.~S.~J.~Robles-P\'erez
informed us another approach to the Wheeler--DeWitt equation \cite{Perez2}.
The massive Klein--Gordon field in curved superspace considered in
Ref.~\cite{Perez2} seems to be equivalent to the dimensional reduction (rather
than the Kaluza--Klein compactification) of the field in the extended superspace.



\acknowledgments
We would like to thank Prof.~S.~J.~Robles-P\'erez and Prof.~P.~V.~Moniz
for providing information about their remarkable works.

\bibliographystyle{apsrev4-1}


\end{document}